\documentclass{mn2e}
\usepackage{epsfig}

\newcommand{\mnref}[1]{\hangindent=0.5in \hangafter=1 #1 \par}
\newenvironment{refs}{\parindent=0pt}{\parindent=1.5em}
\newcommand{\mn}{MNRAS}

\newcommand{\apj}{ApJ}
\newcommand{\apjs}{ApJS}
\newcommand{\aaa}{A\&A}
\newcommand{\aap}{A\&A}
\newcommand{\aas}{A\&AS}

\def\gs{\mathrel{\raise1.16pt\hbox{$>$}\kern-7.0pt
\lower3.06pt\hbox{{$\scriptstyle \sim$}}}}
\def\ls{\mathrel{\raise1.16pt\hbox{$<$}\kern-7.0pt
\lower3.06pt\hbox{{$\scriptstyle \sim$}}}}

\title{Spectropolarimetry of Compton-thin Seyfert 2s}
\author[S.L. Lumsden, D.M. Alexander, J.H. Hough]
{S.L. Lumsden$^{1}$, D.M. Alexander$^{2}$
and J.H. Hough$^{3}$\\
{}$^1$ {\em Department of Physics and Astronomy,
University of Leeds, Leeds LS2 9JT, UK}\\
{}$^2$ {\em Institute of Astronomy, University of Cambridge, Madingley Road,
 Cambridge, CB3 0HA, UK}\\
{}$^3$ {\em Department of Physical Sciences, University of Hertfordshire,
Hatfield, Hertfordshire, AL10 9AB, UK}\\
{Email -- sll@ast.leeds.ac.uk, dma@ast.cam.ac.uk, jhh@star.herts.ac.uk }\\
}

\begin{document}

\label{firstpage}

\maketitle

\begin{abstract}
We present new spectropolarimetry of a sample of nearby Compton-thin Seyfert 2
galaxies (ie those with $N_H<10^{23}$cm$^{-2}$).  We show that the detection
rate of scattered broad H$\alpha$ in this sample is considerably higher than in
Seyfert 2 galaxies as a whole.  Our results also show that in this low
obscuration set it is possible to find scattered broad H$\alpha$ even when the
global properties of the galaxy are largely dominated by the host galaxy and
not the active galactic nucleus.  These results argue against the existence of
a population of `pure' Seyfert 2 galaxies.
\end{abstract}


\section{Introduction}
It is now well established that many Seyfert 2 galaxies show characteristics
consistent with an embedded Seyfert 1 core (ie the presence of a broad line
region) that is hidden from our direct line of sight by surrounding dust.  This
is in good agreement with the basic unified model for Seyfert galaxies
(Antonucci 1993) which suggests that an optically thick circumnuclear
torus, with a scale height at least as large as the broad line region itself,
surrounds the active nucleus.  The orientation of this torus relative to our
line of sight then determines whether the galaxy is classified as a Seyfert 1
or 2 galaxy.

The detection of scattered broad H$\alpha$ lines provides the firmest evidence
for a hidden broad line region (HBLR) in Seyfert 2 galaxies, as has been
demonstrated by, for example, Antonucci and Miller (1985), Miller and Goodrich
(1990), Young et al.\ (1996a), Heisler, Lumsden \& Bailey (1997), Moran et al.\
(2000), Lumsden et al.\ (2001) and Tran (2001, 2003).  A significant fraction
of all Seyfert 2 galaxies show evidence for an HBLR (eg.\ Tran 2001, Moran et
al.\ 2000, Lumsden et al.\ 2001).  This fraction increases as the luminosity of
the active galactic nucleus (AGN) increases (Lumsden \& Alexander 2001, Gu \&
Huang 2002, Martocchia \& Matt 2002, Tran 2003).  It is also clear that the
galaxies known to have HBLRs tend to have warmer mid-far infrared colours than
those galaxies without (Heisler et al.\ 1997, Lumsden et al.\ 2001 and Tran
2001, 2003).

Hard x-ray spectroscopy also allows us to investigate the cores of these
galaxies even at levels of obscuration that would completely hide the broad
line region in the optical (eg Maiolino et al.\ 1998, Risaliti, Maiolino
\& Salvati 1999).
Alexander (2001) pointed out that the combination of the x-ray data and the
spectropolarimetry ruled out the model proposed in Heisler et al.\ (1997) in
order to explain the warmth of the infrared colours in the HBLRs.  Heisler et
al.\ suggested that the colour was simply due to the orientation of the torus
and hence the obscuration to the infrared emitting zone, so that more obscured
systems appeared colder.  By contrast, Alexander showed that the level of x-ray
obscuration did not vary on average between galaxies both with and without
HBLRs.  It has also been shown that the Seyfert 2s which do not
show HBLRs tend to have global properties that are more in keeping with those
of their host galaxies (eg.\ Alexander 2001, Lumsden et al.\ 2001, Tran 2001,
2003).

However, radically different interpretations have been reached from these data.
Tran (2001, 2003) suggests that at least some of the non-HBLRs genuinely lack a
Seyfert 1 core.  Nicastro, Martocchia \& Matt (2003) have suggested that the
main difference between those Seyfert 2s showing scattered broad H$\alpha$ and
those without is the accretion rate of the central AGN.  They suggest that the
rate is simply too low in the Seyfert 2s without scattered broad H$\alpha$ to
support an extensive broad line region.  They note that at least some of these
galaxies have low inferred hard x-ray absorption, suggesting that we should be
able to see the broad line region if it exists.  The Nicastro et al.\ model is
clearly compatible with Tran's observations, though the fraction of galaxies
without an intrinsic broad line region should be less for suitable luminosity
selected samples in the Nicastro et al.\ model than Tran claims to find.  By
contrast, Lumsden \& Alexander (2001) showed that the luminosity dependence of
HBLR detectability was dominant (ie suggesting more luminous active galaxies
supported a larger scattering volume).  Lumsden et al.\ (2001) further
suggested that the level of obscuration was still a weak contribution to the
appearance of AGN once the effects of the intrinsic AGN luminosity were allowed
for, although other studies have not confirmed this result (Alexander 2001,
Tran 2001, Gu et al.\ 2001, Gu \& Huang 2002).  These observations are
compatible with the alternative version of the unified model given by Lawrence
(1991).  He suggested that the inner surface of the torus would be pushed
outwards from the AGN as the luminosity increased purely because the dust
sublimation radius would increase.  In practice an accretion disk has a self
shielding effect so the result may not be as dramtic as Lawrence suggested, but
it is still a potential factor.  It should also be noted that the findings of
Lumsden \& Alexander are not inconsistent with the Nicastro et al.\ result,
since the accretion rate may be the fundamental parameter determining the AGN
luminosity, and the presence of suitable scatterers must scale with the
luminosity of the source.

The aim of this paper is to test which of these possibilities is a better match
to the actual results for Seyfert 2s with known low obscuration from hard x-ray
data.  In these sources obscuration to the scattering sites should be
effectively irrelevant if the extinction is reasonably compact.  This should
allow us to test the other models outlined above more easily than in previous
samples which span the full range of obscuration.  In order to test these ideas
we have acquired spectropolarimetry at both high signal-to-noise and higher
spectral resolution than in our previous observations.  The latter allows a
better separation of possible multiple polarising components.

\section{Observations and Data Reduction}
Our observations were acquired on the nights of 1--4 June 2002 at the
Anglo-Australian Telescope.  We used the RGO Spectrograph with an EEV
4096$\times$2048 pixel CCD.  Vignetting within the instrument limits the actual
spectral coverage to 3296 pixels.  The slit width was typically 1.5 arcseconds,
matched to the approximate seeing.  The effective spectral resolution of our
data is $\sim3$\AA.  Conditions were photometric on 3 and 4 June, and partly
photometric on 2 June.  Where fluxes are reported they are taken from the 
photometric data sets alone.  Most sources were observed on more than one night
because of the requirement for high signal-to-noise data.  All were observed
on at least one of the photometric nights.

A calcite prism was used to split the incoming beam into $e$ and $o$-rays
together with an aperture mask made up of discrete slit-lets.  A half-waveplate
modulated the incoming phase.  Four steps of the waveplate at 0$^\circ$,
45$^\circ$, 22.5$^\circ$ and 67.5$^\circ$ were required to derive the full set
of Stokes parameters for linearly polarized light.  Sky subtraction was
achieved by nodding the object into an adjacent slit-let.  The sky and object
spectra for the four separate waveplate positions were extracted then combined
to give the final Q, U and I Stokes parameters.  We extracted data from a
region of between 3 and 4 arcseconds along the slit in most cases.

We also obtained observations of polarised and unpolarised standard stars in
order to check the system efficiency and calibrate the position angle data.
The results showed good agreement with published values for the polarised
standards, and confirmed that the system polarisation was less than 0.1\% from
the unpolarised standards.

We selected our targets from the catalogue of Seyfert 2s with previously
available x-ray spectroscopy given in Bassani et al.\ (1999).  We imposed only
two criteria: that the object was sufficiently southern to be observable from
the AAT, and that it had an inferred neutral hydrogen column density
$<10^{23}$cm$^{-2}$.  We did include two other targets in the sample as well to
fill gaps in the RA distribution.  These were IC5063, previously known to show
broad scattered H$\alpha$ but for which spectropolarimetry of the H$\beta$
region does not exist, and NGC6300, which is not listed in Bassani et al.\ (see
Risaliti 2002 for x-ray data).  Both of these have
$N_H\sim2\times10^{23}$cm$^{-2}$.  We integrated on source until we had a clear
detection of an HBLR or for approximately 8 hours.  The final signal-to-noise
is similar to the data presented in Lumsden et al.\ (2001), though at an
effective spectral resolution that is 2.5 times higher.  The full list of all
seven sources, including the individual integration times, is given in Table 1.

All of our targets are relatively bright x-ray sources, and most of the low
column density targets fall within the sample of narrow line x-ray galaxies
identified in the sample of Ulvestad \& Wilson (1989).  These observations
therefore provide a useful comparison sample to that of Moran et al.\ (2000),
since they obtained spectropolarimetry of the classical Seyfert 2s in the
Ulvestad \& Wilson sample, but not the narrow line x-ray galaxies.  In practice
some of our targets have previously been classified as Seyfert 1.8 or 1.9.  We
opted to observe these sources as it is not clear whether the weak broad
component seen in direct light is scattered in any event.  We therefore count
all such classification as Seyfert 2s.

The data were reduced in a standard fashion to give Q, U and I Stokes
parameters.  Since the normal definition of polarisation is a positive definite
quantity, we prefer to work with the rotated Stokes parameter $Q'$.  This is
the result of rotating the Stokes parameters through an angle consistent with
the measured position angle.  The net result is that virtually all of the
significant polarisation information is rotated into the new $Q'$ parameter.
Of course this is only strictly valid when the position angle is approximately
constant with wavelength.  This is largely true for our data, since the
observed variation with wavelength is always small.  The actual observed
mean position angle is given in Table 1.

Line fluxes were measured from both the total intensity and polarised flux
data.  The fluxes are taken from Gaussian line fits.  All narrow lines required
at least two component fits to match the resolved line structure.  Where
necessary we also fitted a component to broad H$\alpha$ in the direct light
spectrum.  There was no need to fit a component to broad H$\beta$ for any
object in the direct light spectrum.  The fitted components were constrained so
that, for example, the wavelengths of the line centres of the two [NII] lines
around H$\alpha$, and their relative intensities, matched theoretical
expectation.  We only fitted a single Gaussian where a broad component to
H$\alpha$ is clearly present in the polarised light spectra, since this
adequately accounts for all of the flux and there is no evidence for a strong
polarised narrow line component (as evidenced in Figure 1 by the generally weak
or absent [OIII] 5007\AA\ line in polarised light).  The exceptions to this are
NGC~2992 and F18325--5926 which are discussed further in Section 3.  In these
cases we allowed for the narrow line contributions as well.  Where broad
H$\alpha$ was not present in the polarised flux, we estimated limits to the
broad H$\alpha$ line assuming a line width of 3500kms$^{-1}$ (the mean from the
sample of Seyfert 1s in Stirpe 1990).  The resultant observed
scattered broad H$\alpha$ fluxes are
given in Table 1.  Our deepest limit, for NGC~5506, is approximately a factor
of two lower than for any of the non-detections reported in Lumsden et al.\
(2001).  It is worth noting however that all of the detections we report here
have broad H$\alpha$ fluxes that lie above the limits we found for the
non-detections in Lumsden et al.\ (2001), so it is fair to compare the results
of this survey with the earlier work.  The key results are given in Table 1
with the observed (not extinction corrected) fluxes for [OIII] 5007\AA\ and the
broad H$\alpha$ component seen in scattered light.

\section{Results}
The basic results of our observations are shown in Figure 1.  This Figure shows
the direct light spectrum, $I$, the rotated Stokes parameter, Q$'$ and the
percentage polarisation.  Three of our seven sources (NGC~7314, MCG-5-23-16 and
IC~5063) show clear evidence for broad H$\alpha$ in the polarised flux.  Of
these, NGC~7314 and MCG-5-23-16 were not previously known to contain HBLRs.
IC~5063 was previously found to contain a HBLR by Inglis et al.\ (1995), and
that result is confirmed here.  None show any evidence for scattered H$\beta$
however.

The nature of the other four sources we observed required further analysis.
There are clear features present in the polarisation for NGC~2992 and
F18325--5926, though the polarised flux is dominated by emission from the
narrow line region.  It is possible to decouple different polarisation
mechanisms where there is evidence for a change in polarisation with
wavelength.  For example, if the bulk of the polarisation observed is due to
dichroic absorption in the host galaxy, but any broad lines present are
scattered, there is no {\em a priori} reason for the position angle, or the
degree of polarisation, of the two mechanisms to be the same.  We can subtract
off the smooth component (whether due to dichroic absorption or large scale
scattering is irrelevant) to leave only the polarisation due to the scattering
from the broad line region.  In practice, we fit a low-order polynomial to the
$Q'/I$ data to derive the smooth component.  The residual is then rescaled by
the total intensity to give a polarised flux spectrum as shown in Figure 2.

We applied this procedure to the four sources that did not show immediate
evidence for broad H$\alpha$ to determine if there was evidence for a masked
scattered component.  Both NGC~2992 and F18325--5926 exhibit behaviour of this
kind, as indicated by their polarisation spectra.  The resultant polarised flux
for these objects after removal of the smooth component is shown in Figure 2,
revealing clear broad H$\alpha$.  Again this is the first detection of
scattered broad H$\alpha$ in these galaxies.  Both also show some evidence for
broad H$\beta$ as well.  The remaining two sources, NGC~5506 and NGC~6300, do
not show any evidence for broad H$\alpha$ however.  We have given limits to the
broad H$\alpha$ line as noted in Section 2.  Overall we detect a clear
signature of scattered broad H$\alpha$ in 5 out of 7 sources.


It is worth briefly considering the non-detections, NGC~5506 and NGC~6300.
NGC~5506 is the more interesting case.  Nagar et al.\ (2002) present compelling
evidence that NGC~5506 is an obscured narrow line Seyfert 1 based on near
infrared spectroscopy.  In keeping with previous papers by us however we do not
class this object as an HBLR, which we define strictly as those galaxies
showing polarised broad H$\alpha$.  It clearly has a very low column density to
the x-ray source (Risaliti 2002).  However, NGC~5506 is almost edge-on, and has
a conspicuous dust lane, so the obscuration is on considerably larger scales
than in the other galaxies we have studied.  The polarised flux is consistent
with a dichroic origin, from dust in front of the narrow line region.  There is
certainly no evidence that we can see the narrow line Seyfert core, since the
intrinsic line width found by Nagar et al.\ (and Veilleux, Goodrich \& Hill
1997) in the near infrared is considerably broader than the optical lines.
Equally, there is no reason to suspect that a narrow line Seyfert 1 should not
support an extensive scattering region if it is luminous enough.  Therefore the
most likely cause of the non-detection is the extent of the obscuration
rather than the actual amount.  We can say less about NGC~6300, since it is the
least luminous galaxy and although the limits we can place on broad H$\alpha$
are strict, they are not inconsistent with previous limits for galaxies of this
luminosity.  Therefore we cannot say whether the source has an HBLR that is too
faint to detect with the current instrumentation or lacks an HBLR completely.

\section{Comparison with previous surveys}
\subsection{Overall detection rates}
Our new data clearly indicate that it is possible to detect HBLRs in most low
obscuration Seyfert 2s.  We can extend this result slightly by considering all
of the low obscuration Seyfert 2s in the Bassani et al.\ (1999) sample.  There
are another six galaxies with $N_H<10^{23}$cm$^{-2}$ for which
spectropolarimetric data exists.  Four show HBLRs (05189--2524, NGC~5252,
20460+1925, 23060+0505: Young et al.\ 1996a,b), the other two do not (NGC~7172,
NGC~7590: Lumsden et al.\ 2001, Alexander et al.\ in preparation).  It should
be stressed however that the signal-to-noise of the polarimetry for the latter
two objects is not as good as for NGC~5506 for example.  Despite this, and
although these samples are not complete, the indication is that the detection
rate of HBLRs in low obscuration Seyfert 2s is higher than in the population as
a whole (eg.\ see the overall detection rate as given in Lumsden et al.\ 2001,
Tran 2001 or Moran et al.\ 2000).  This ties in with the previously more
tentative suggestion in Lumsden et al.\ (2001) that it is easier to find HBLRs
at low column density.  However, similar high spectral resolution observations
of a suitable sample of Compton-thick Seyfert 2s is required to test
this directly.

We can also compare our results on the Seyfert 1.8/1.9 galaxies in our sample
with those of Goodrich (1989).  Noteably, all of the galaxies that would be
classified as a Seyfert 1.9 by us also show an HBLR.  Goodrich (1989) by
comparison found that only a small fraction of his targets showed evidence of
an HBLR.  It is not clear if this difference is due to our selection being
amongst the brighter members of this class (which maybe why they fall into the
narrow line x-ray galaxy class of Ulvestad \& Wilson), or to poorer
signal-to-noise in Goodrich's data.  Extra studies would certainly be
worthwhile to determine if HBLRs are common characteristics of other members of
this class.

\subsection{Infrared, radio and x-ray diagnostics}
We have not yet considered the other properties of these galaxies, and how they
fit into any overall picture.  In previous papers we showed that the other key
factors in determining HBLR visibility were the core luminosity of the AGN and
the contrast of the AGN and host galaxy luminosities (and, perhaps,
obscuration).  In order to avoid problems with varying signal-to-noise data we
only consider the galaxies with spectropolarimetry presented in this paper.
Unfortunately, MCG-5-23-16 lies in the IRAS gap region, and therefore has no
far infrared data available, so it is not included.

The mid-to-far infrared colours were shown in both Heisler et al.\ (1997) and
Lumsden et al.\ (2001) to be a good discriminator between galaxies with HBLRs
and those without, and also a good indicator of the ratio of host galaxy and
nuclear luminosities (Alexander 2001).  In particular, previously known HBLRs
have IRAS colours similar to those of a reddened Seyfert 1, whereas most
non-HBLRs have colours that are very similar to a star forming galaxy.  Another
good discriminator as to the activity present is the ratio of low frequency
radio data with the far-infrared flux (Lumsden et al.\ 2001 and Tran 2001,
2003).  We used 20cm NVSS data (Condon et al.\ 1998), which has the drawback of
sampling much of the emission from the host galaxy as well as the AGN.
However, the radio data should be dominated by the AGN except at the lowest
luminosities.  This is clearly true for the sample considered here since the
comparison of radio and absorption corrected hard x-ray luminosities shows a
scatter of less than 30\%.

The data for the galaxies presented here gives a somewhat different picture
to that presented by Tran (2001, 2003:  Figure 3, 
and compare with similar figures in Lumsden et al.\ 2001).  Two of
the HBLRs have infrared colours consistent with star forming galaxies (NGC~2992
and NGC~7314), and one of the non-HBLRs has the colour of a reddened Seyfert 1
(NGC~5506).  This result indicates that it is possible to find relatively low
luminosity Seyfert 1 cores in galaxies whose global properties are dominated by
their host galaxies rather than the nucleus.  In particular, we stress that
some of the HBLRs in this sample have flux ratios $F_{60}/F_{25}>4$, cooler
than any found previously.  The same is true for the far-infrared/radio
comparison.  Again, the HBLRs are not confined to the AGN dominated region.
Tran (2001, 2003) made much of the absence of HBLRs from the lower left of his
equivalent of Figure 3(b) (which in our plot is actually the lower right).
This argues strongly against the simple split between a group of Seyfert 2s
without an HBLR and those with as Tran proposes.

Nicastro et al.\ (2003) have recently suggested that the difference between a
detection and a non-detection may actually lie in the accretion rate onto the
central black hole.  The accretion rate can be constrained by the hard x-ray
flux, and the mass of the central black hole by the extinction corrected bulge
luminosity (though with larger scatter).  We have applied their technique for
estimating the accretion rate to our sample where suitable data exists.  The
results are shown in Table 1.  Our results indicate that all of our sources,
HBLRs and non-HBLRs alike, lie above the minimum accretion rate of $10^{-3}$ of
the Eddington rate suggested by Nicastro et al.  This sample may not be the
best to test the Nicastro et al.\ suggestion however, since all of the Seyfert
2s considered here, apart from NGC~6300, are relatively luminous compared to
the whole local AGN population, and X-ray selected AGN as a class are also
unlikely to have a low accretion rate. We discuss a better test of the Nicastro
et al.\ results in Lumsden \& Alexander (2003, in preparation).

\section{Discussion}
Our new results on Compton thin Seyfert 2 galaxies indicate that most show
evidence for an HBLR.  The exceptions are NGC~5506, which has very extended
obscuration but is known from infrared spectroscopy to have a narrow line
Seyfert 1 core, and NGC~6300 for which we do not have sufficient
signal-to-noise to definitely state whether an HBLR is present or not.  At
least one factor in our success rate in detecting HBLRs is the higher spectral
resolution data we have acquired.  This makes it much simpler to detect weak
broad scattered lines masked by polarised narrow lines as shown in Section 3.
Our results also show counter examples of HBLRs in the `pure' Seyfert 2 region
of radio/infrared colour space proposed by Tran (2001, 2003), and for the first
time we find HBLRs in galaxies with cool far-infrared colours
($F_{60}/F_{25}>4$).  This strongly suggests that the split proposed by Tran is
not as simple as appears from his data.

It is still worth considering briefly the evidence for a class of Seyfert 2s
lacking a broad line region.  Nicastro et al.\ (2003) have shown that at least
some of the `pure' Seyfert 2s identified by Tran must have very low accretion
rates.  However, some caution must be applied.  X-ray and spectropolarimetric
data are rarely contemporaneous, and it is known that there can be wide
variations in the observed level of x-ray flux with time (eg Turner et al.\
1997).  A good example to consider here is the case of NGC~2992, which has
varied in x-ray luminosity by over an order of magnitude in the last twenty
years (Gilli et al.\ 2000), reaching a maximum again recently.  Our
spectropolarimetry coincides with this rise to maximum.  At minimum, NGC~2992
would have fallen well below the minimum accretion cut-off suggested by
Nicastro et al.  The narrow line region responds much less to such variability
in the central engine since it is spatially much larger, and hence the lag
between the inner and outer edge tends to smooth out variations.  Longer
timescale monitoring of the x-ray properties of the low accretion rate Seyfert
2s would be useful in determining if their lack of an HBLR is due to some
intrinsic difference in the Seyfert 2 population as a whole (which we would
argue against) or simply that they are in a low state.  Once accretion stops
completely such objects would rapidly lose their narrow line regions as well,
and cease to classed as Seyfert 2s.

\section{Acknowledgments}

SLL acknowledges the support of PPARC through the award of an Advanced Research
Fellowship.  DMA acknowledges the Royal Society for the generous support
provided by a University Research Fellowship.  We thank the staff of the AAO
for their help in acquiring these observations, and the referee for somments
that have improved the text.

\parindent=0pt

\vspace*{3mm}

\section*{References}
\begin{refs}
\mnref{Alexander, D.M., 2001, \mn, 320, L15}
\mnref{Antonucci, R., Miller, J.S., 1985, ApJ, 297, 621}
\mnref{Antonucci, R. 1993, ARA\&A, 31, 473}
\mnref{Bassani, L., Dadina, M., Maiolino, R., Salvati, M., Risaliti, G.,
        della Ceca, R., Matt, G., Zamorani, G., 1999, \apjs,  121, 473}

\mnref{Condon, J.J., Cotton, W.D., Greisen, E.W., Yin, Q.F., Perley, R.A., 
  Taylor, G.B., Broderick, J.J., 1998, AJ, 115, 1693}
\mnref{Gilli, R., Maiolino, R., Marconi, A., Risaliti, G., Dadina, M., 
  Weaver, K.A., Colbert, E.J.M., 2000, \aap, 355, 485 }
\mnref{Goodrich, R.W.\ 1989, \apj, 340, 190 }
\mnref{Gu, Q., Maiolino, R., Dultzin-Hacyan, D., 2001, \aaa, 366, 765}
\mnref{Gu, Q., Huang, J., 2002, \apj, 579, 205 }
\mnref{Heisler, C.A., Lumsden, S.L., Bailey, J.A., 1997, Nature, 385, 700}
\mnref{Inglis, M.D., Young, S., Hough, J.H., Gledhill, T., Axon, D.J., Bailey,
        J.A., Ward, M.J., 1995, \mn, 275, 398}

\mnref{Lawrence, A., 1991, \mn,  252, 586 }

\mnref{Lumsden, S.L., Alexander, D.M., 2001, \mn, 328, L32 }
\mnref{Lumsden, S.L., Heisler, C.A., Bailey, J.A., Hough, J.H., Young, S.,
        2001, \mn, 327, 459}
\mnref{Maiolino, R., Salvati, M., Bassani, L., Dadina, M., della Ceca, R., 
  Matt, G., Risaliti, G., Zamorani, G., 1998, \aaa, 338, 781 }
\mnref{Miller, J.S., Goodrich, R.W., 1990, \apj, 355, 456 }
\mnref{Moran, E.C., Barth, A.J., Kay, L.E., Filippenko, A.V., 2000, \apj,
        540, L73}
\mnref{Nagar, N.M., Oliva, E., Marconi, A., Maiolino, R., 2002, \aaa, 391, L21}
\mnref{Nicastro, F., Martocchia, A., Matt, G., 2003, \apj, 589, L13 }
\mnref{Risaliti, G., Maiolino, R., Salvati, M.\ 1999, \apj, 522, 157}
\mnref{Risaliti, G., 2002, \aaa,  386, 379 }
\mnref{Stirpe, G.M., 1990, \aas, 85, 1049}

\mnref{Tran, H.D, 2001, \apj, 554, L19}
\mnref{Tran, H.D., 2003, \apj, 583, 632}
\mnref{Turner, T.J., George, I.M., Nandra, K., Mushotzky, R.F., 1997, 
        \apjs, 113, 23}
\mnref{Ulvestad, J.S., Wilson, A.S., 1989, \apj, 343, 659}
\mnref{Veilleux, S., Goodrich, R.W., Hill, G.J., 1997, \apj, 477, 631 }
\mnref{Young, S., Hough, J.H., Efstathiou, A., Wills, B.J., Bailey, J.A., Ward,
M.J., Axon, D.J.,  1996a, \mn, 281, 1206}
\mnref{Young, S., Hough, J.H., Axon, D.J., Ward, M.J., Bailey, J.A., 1996b, 
  \mn, 280, 291 }

\end{refs}

\newpage

\onecolumn

\begin{table}

\begin{tabular}{lcrcccccccc}
\multicolumn{1}{c}{Name} &
\multicolumn{1}{c}{cz} &
\multicolumn{1}{c}{Exp. Time} &
\multicolumn{1}{c}{F$_{\lambda5007}$} &
\multicolumn{1}{c}{F$_{\lambda6563b}$} &
\multicolumn{1}{c}{H$\alpha$/H$\beta$} &
\multicolumn{1}{c}{W$_{\lambda5007}$} &
\multicolumn{1}{c}{$L/L_E$}&
\multicolumn{1}{c}{$\theta$}\\

\multicolumn{1}{c}{} & \multicolumn{1}{c}{(kms$^{-1}$)} & 
\multicolumn{1}{c}{(s)} &
\multicolumn{1}{c}{(10$^{-13}$ergs/s/cm$^2$)} &
\multicolumn{1}{c}{(10$^{-15}$ergs/s/cm$^2$)} &
 \multicolumn{1}{c}{} & 
\multicolumn{1}{c}{(\AA)} & $10^{-3}$
& \multicolumn{1}{c}{$^\circ$}\\
 NGC~2992  & 2310 &  9600 & 2.8 &  19.6 &  5.3 &   85 &
 5.1 & 35\\
 MCG-5-23-16 & 2480 &  9600 & 5.6 &  42.6 &  2.0 &    51 & 19\\
 NGC~5506  & 1850 &  33600 & 12.5 & $<$1.0 &  4.9 &   595
 & 7.3 & 80\\
 F18325--5926 & 6060 &  28800 & 10.7 & 22.4 &  3.5 &   15\\
 NGC~7314  & 1420 &  14400 & 2.0 &  5.6  &  4.8 &   164 &
 2.3 & 5\\
					  				  
 NGC~6300  & 1110 &  28800 & 1.4 & $<$1.3  &  8.7 &    36
 & 1.3  & 35\\
 IC~5063 & 3400 &  3360 & 9.5 &  29.6 &  4.3 &  160 & 2.4 & 8\\
\end{tabular}

\caption{Observed properties of our sample.  The exposure time in seconds
refers to the total on source time.  F$_{\lambda5007}$ is the observed [OIII]
5007\AA\ line flux, F$_{\lambda6563b}$ is the observed broad component to the
scattered H$\alpha$ flux (or the limit that can be placed on it) as discussed
in Section 2, H$\alpha$/H$\beta$ is the observed Balmer
decrement from the narrow line component fits.  W$_{\lambda5007}$ is the
observed [OIII] 5007\AA\ equivalent width and is partly a measure of the
intrinsic ratio of AGN and host galaxy.  $L/L_E$ is the predicted ratio of the
nuclear luminosity to the Eddington luminosity derived in the same fashion as
Nicastro et al.\ (2003), and is a measure of the accretion rate.  This column
is left blank if the appropriate data to calculate this quantity do not exist.
}
\end{table}

\vspace*{5in}

\newpage

(a) \vspace*{-1.2cm}

\hspace*{0.1in} \begin{minipage}{6.5in}{
\psfig{file=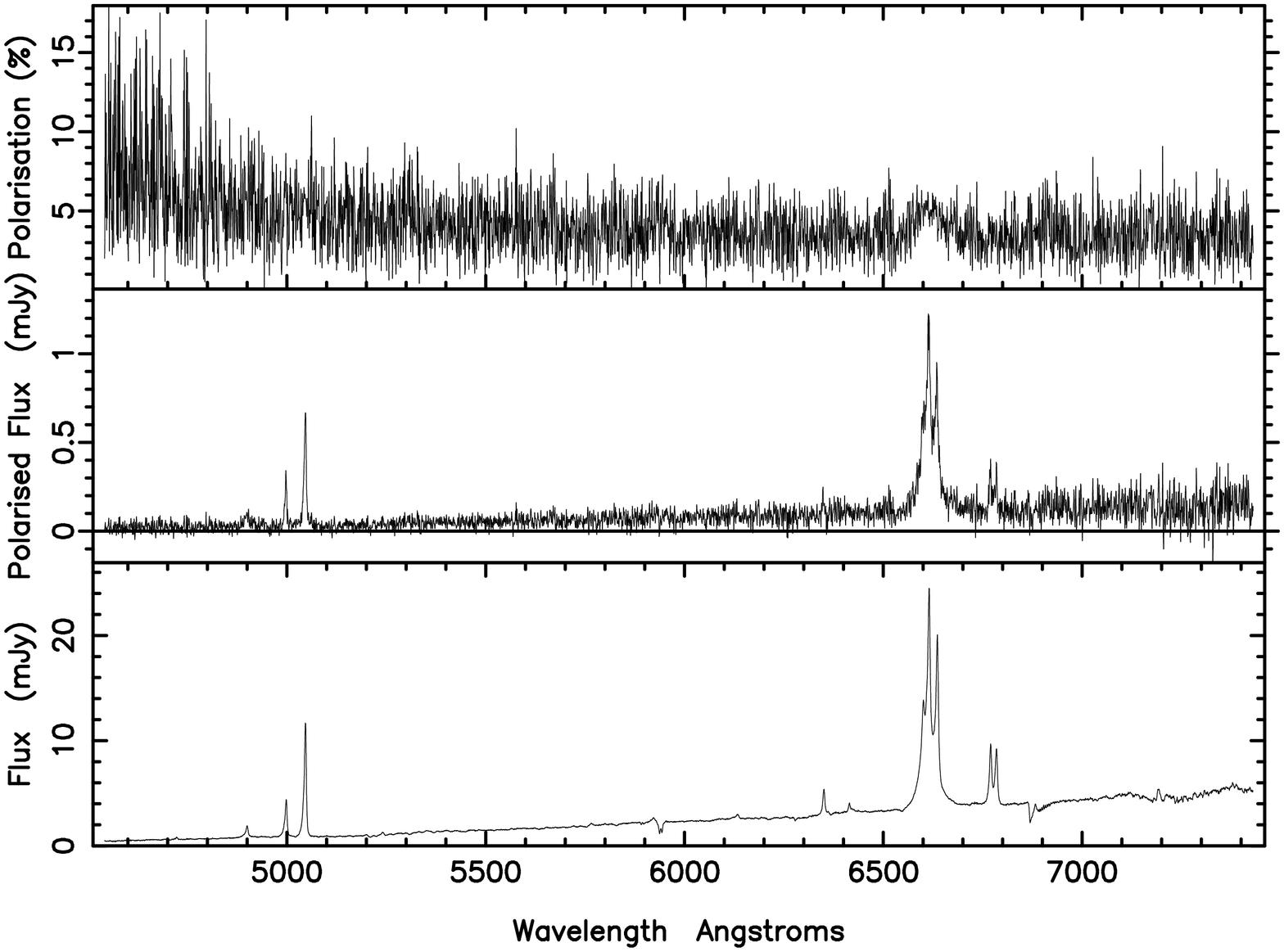,width=4.5in,angle=-90,clip=}}\end{minipage}

(b) \vspace*{-1.2cm}

\hspace*{0.1in} \begin{minipage}{6.5in}{
\psfig{file=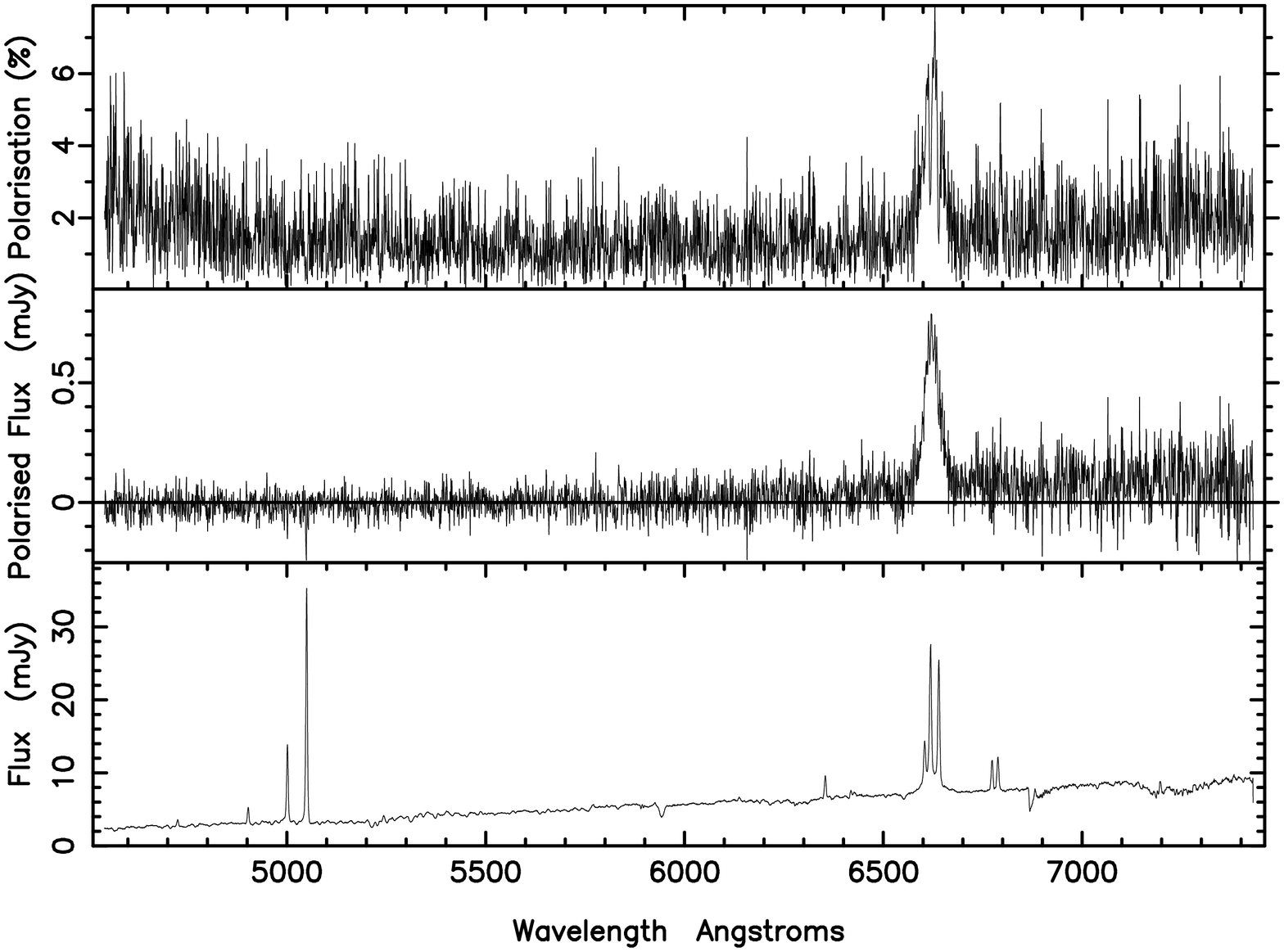,width=4.5in,angle=-90,clip=}}\end{minipage}

\begin{minipage}{\textwidth}{
{\bf Figure 1:} Spectropolarimetric data for our sample, showing in
each case, from top to bottom, the observed position angle, the rotated
Stokes parameter $Q'$ and the total intensity.  The data are unbinned.
The plots are for (a) NGC~2992 and (b) MCG-5-23-16.
}\end{minipage}

(c) \vspace*{-1.2cm}

\hspace*{0.1in} \begin{minipage}{6.5in}{
\psfig{file=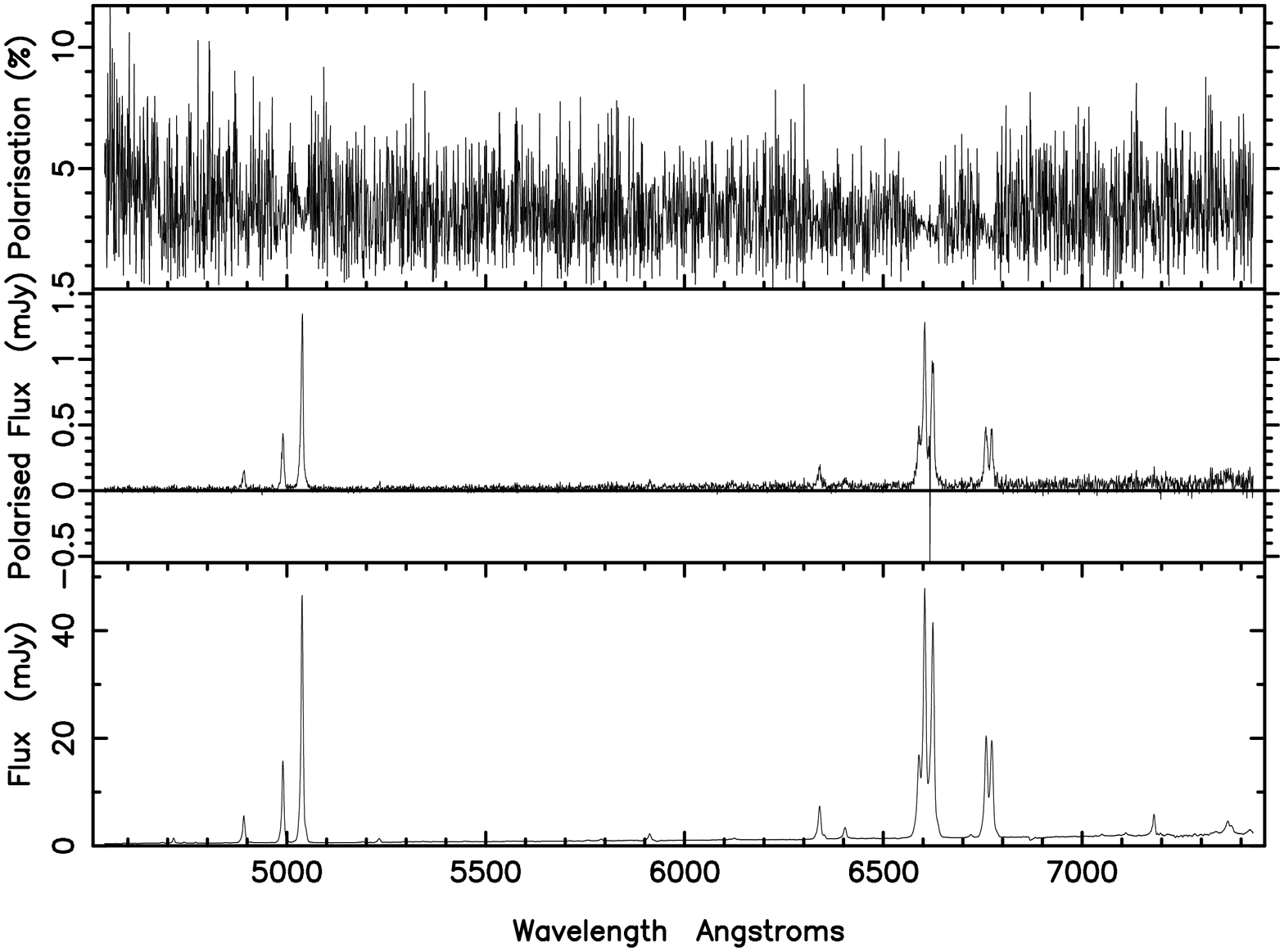,width=4.5in,angle=-90,clip=}}\end{minipage}

(d) \vspace*{-1.2cm}

\hspace*{0.1in} \begin{minipage}{6.5in}{
\psfig{file=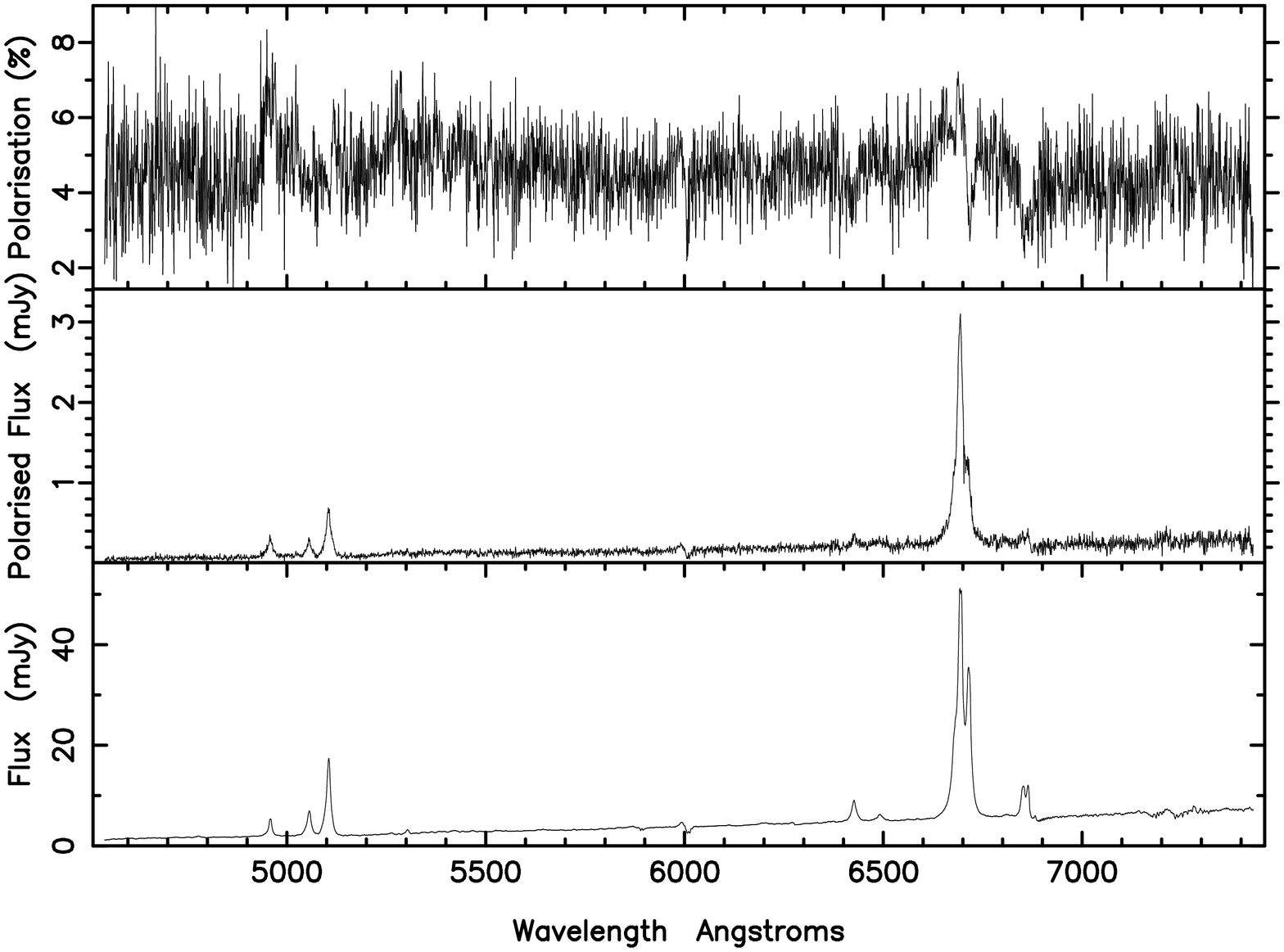,width=4.5in,angle=-90,clip=}}\end{minipage}

\begin{minipage}{\textwidth}{
{\bf Figure 1} continued: (c) NGC~5506 and (d) 18325--5926.
}\end{minipage}

(e) \vspace*{-1.2cm}

\hspace*{0.1in} \begin{minipage}{6.5in}{
\psfig{file=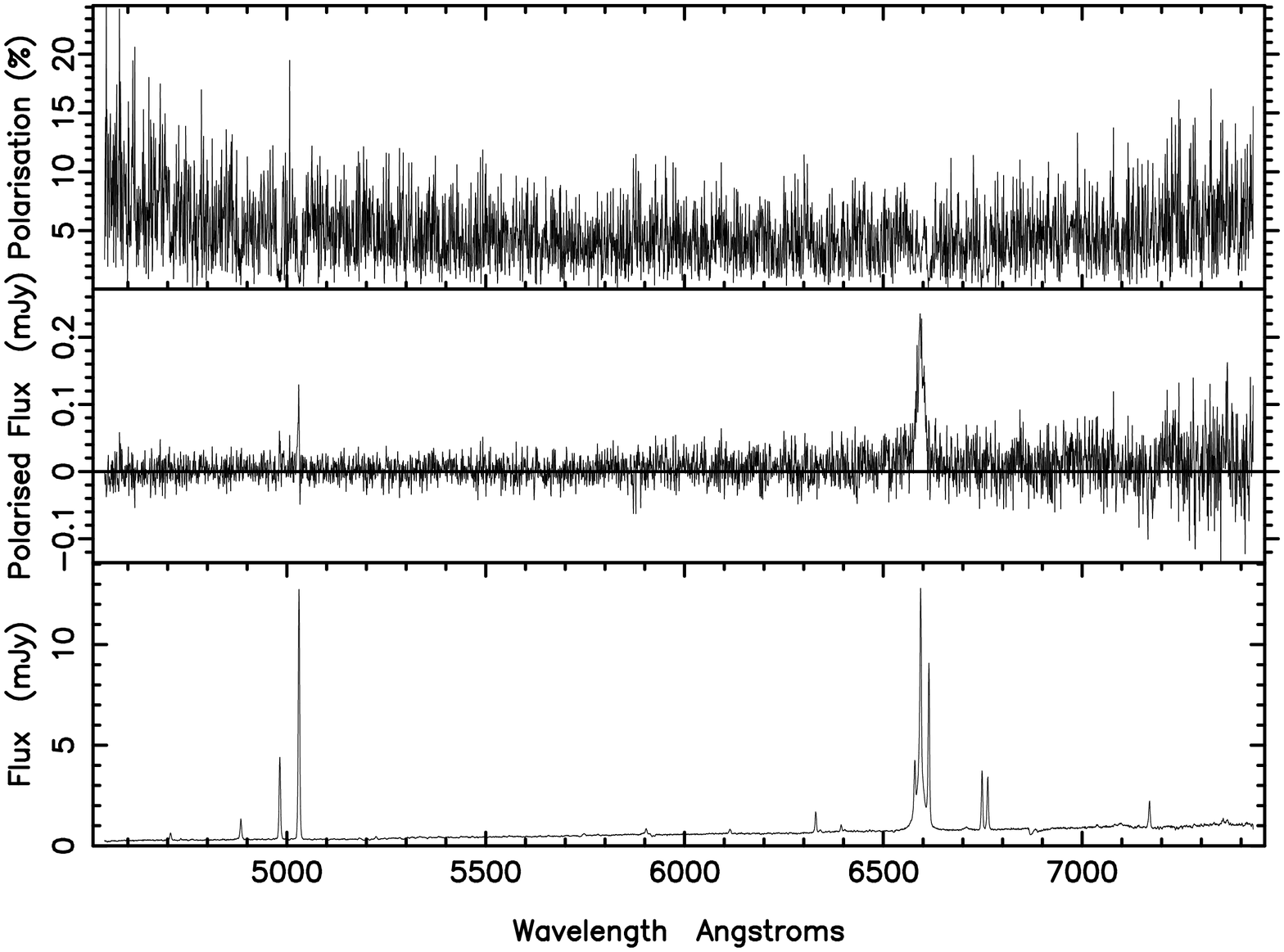,width=4.5in,angle=-90,clip=}}\end{minipage}

(f) \vspace*{-1.2cm}

\hspace*{0.1in} \begin{minipage}{6.5in}{
\psfig{file=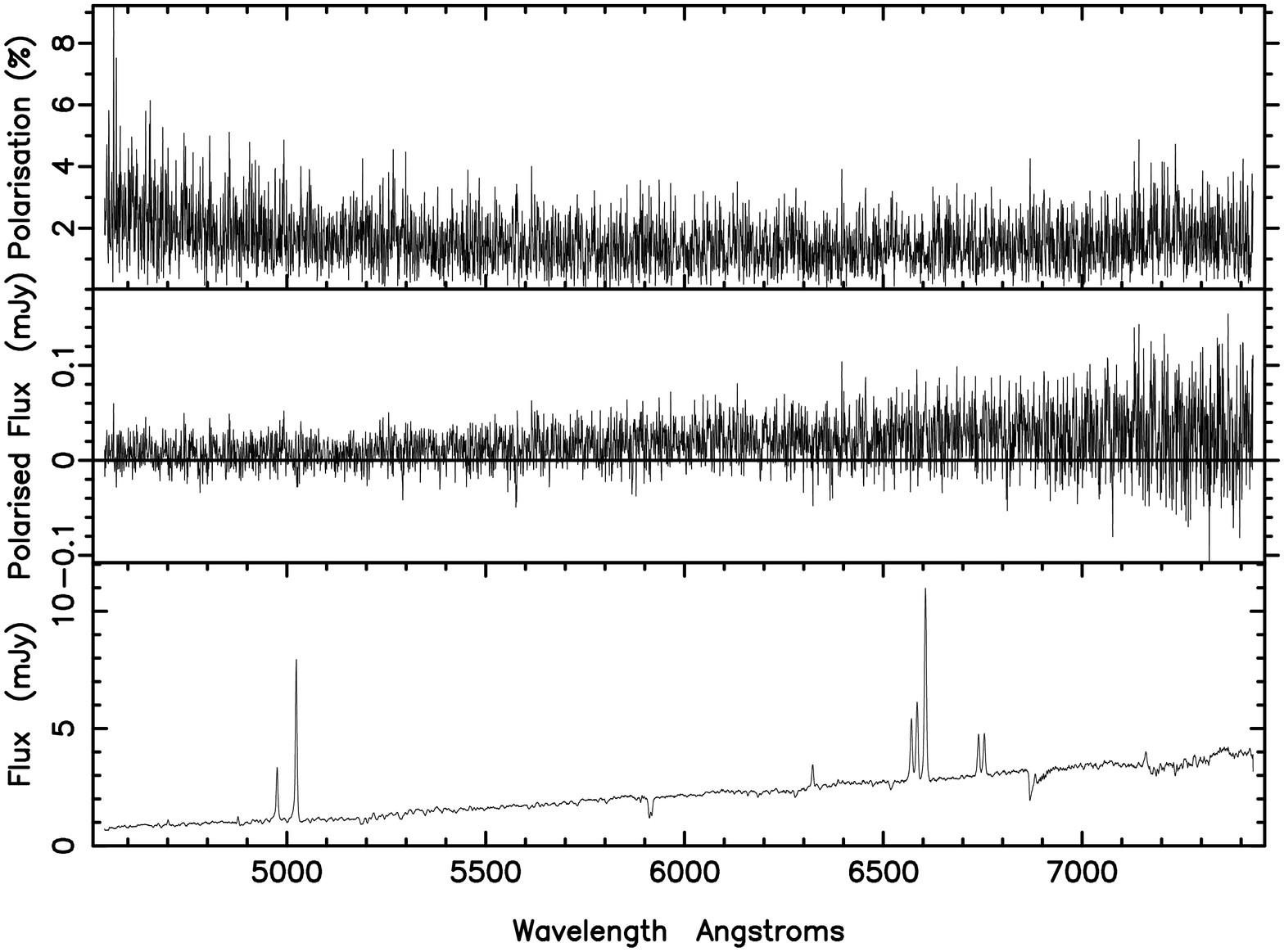,width=4.5in,angle=-90,clip=}}\end{minipage}

\begin{minipage}{\textwidth}{
{\bf Figure 1} continued:  
(e) NGC~7314 and (f) NGC~6300.
}\end{minipage}

(g)  \vspace*{-0.8cm}

\hspace*{0.1in} \begin{minipage}{6.5in}{
\psfig{file=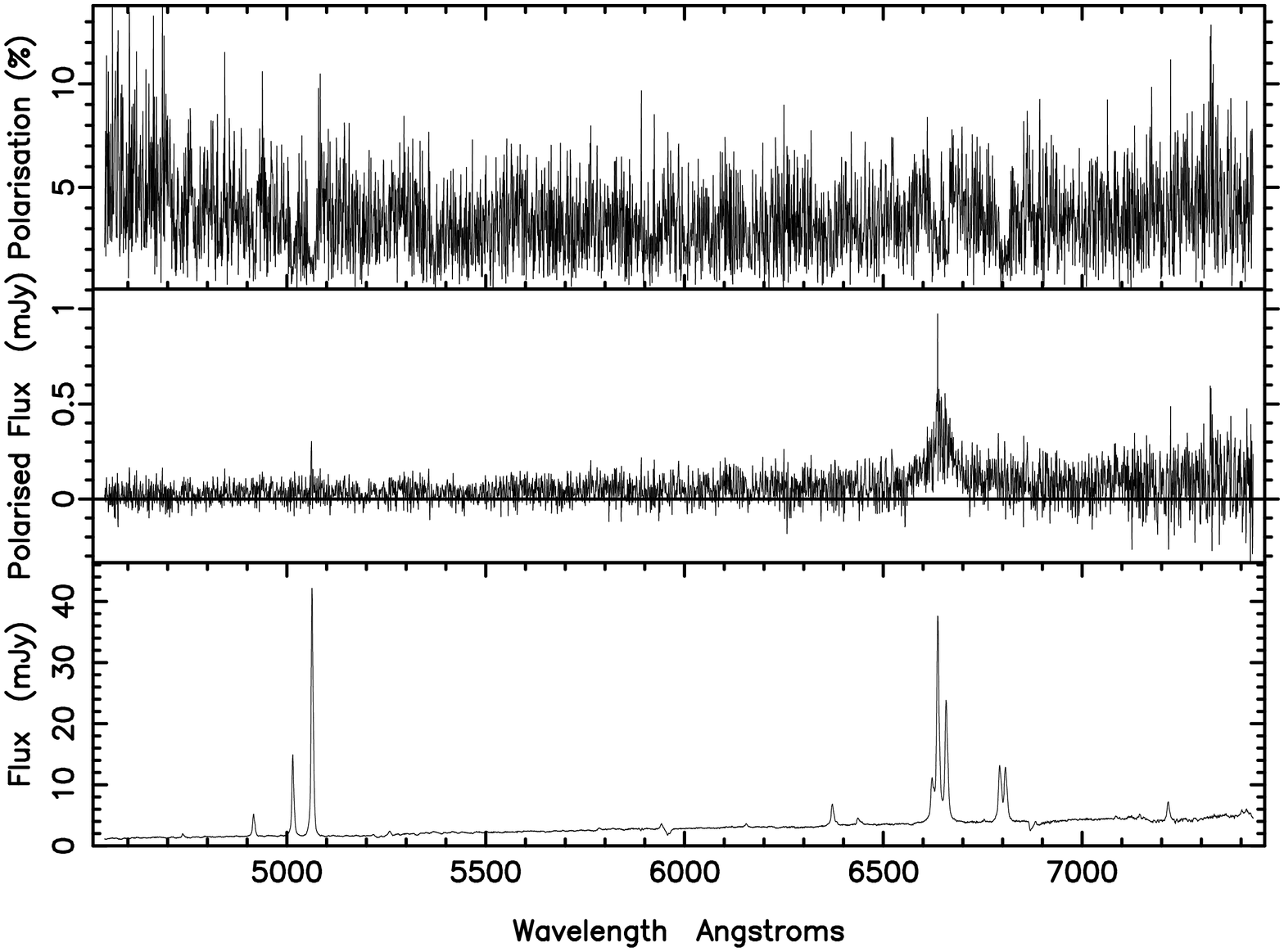,width=4.5in,angle=-90,clip=}}\end{minipage}

\begin{minipage}{\textwidth}{
{\bf Figure 1} continued:  (g) IC~5063
}\end{minipage}

\vfill

\newpage

(a) \vspace*{-1.2cm}

\begin{minipage}{6.5in}{
\psfig{file=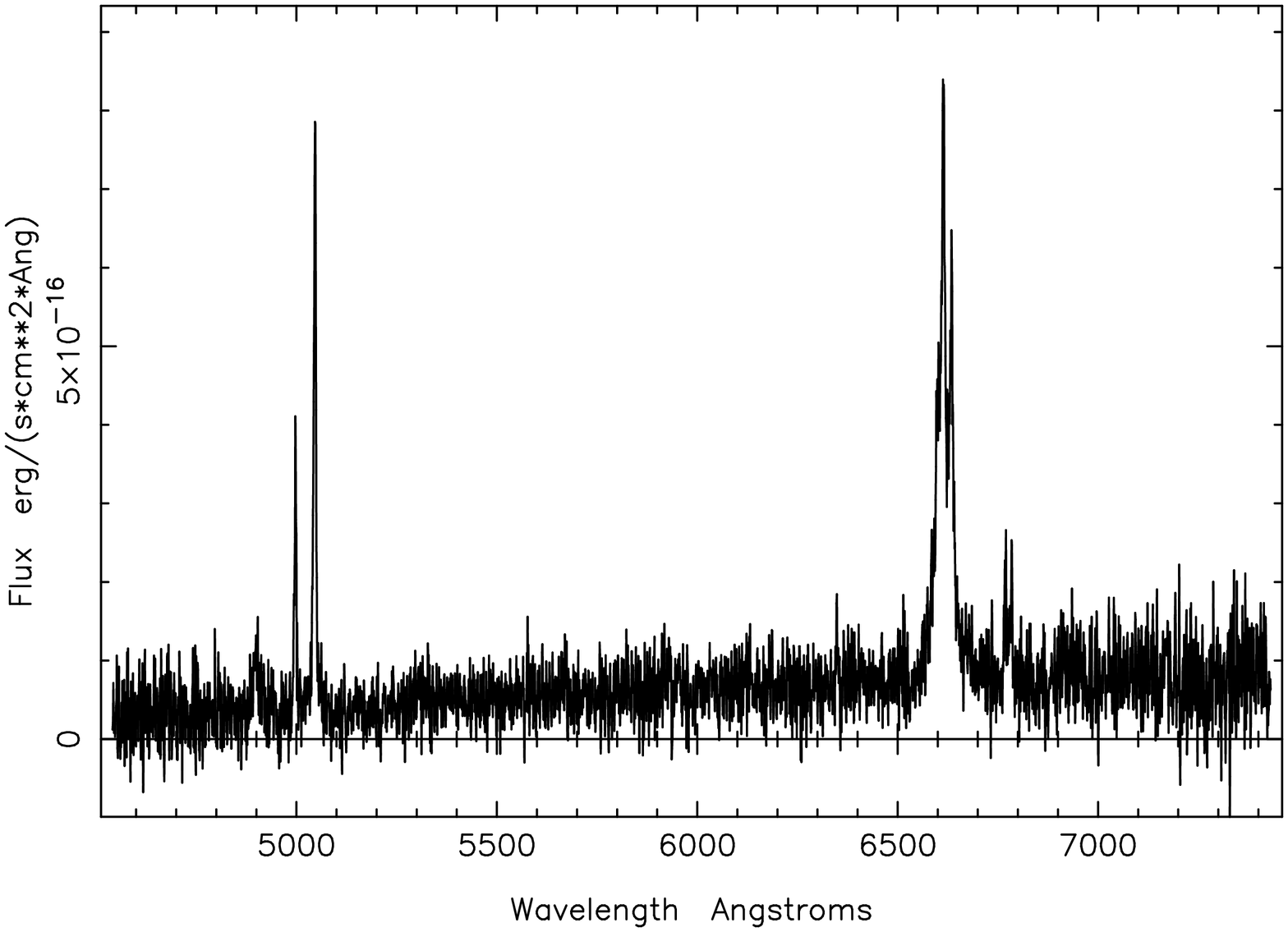,width=4.5in,angle=-90,clip=}}\end{minipage}

(b) \vspace*{-1.2cm}

\begin{minipage}{6.5in}{
\psfig{file=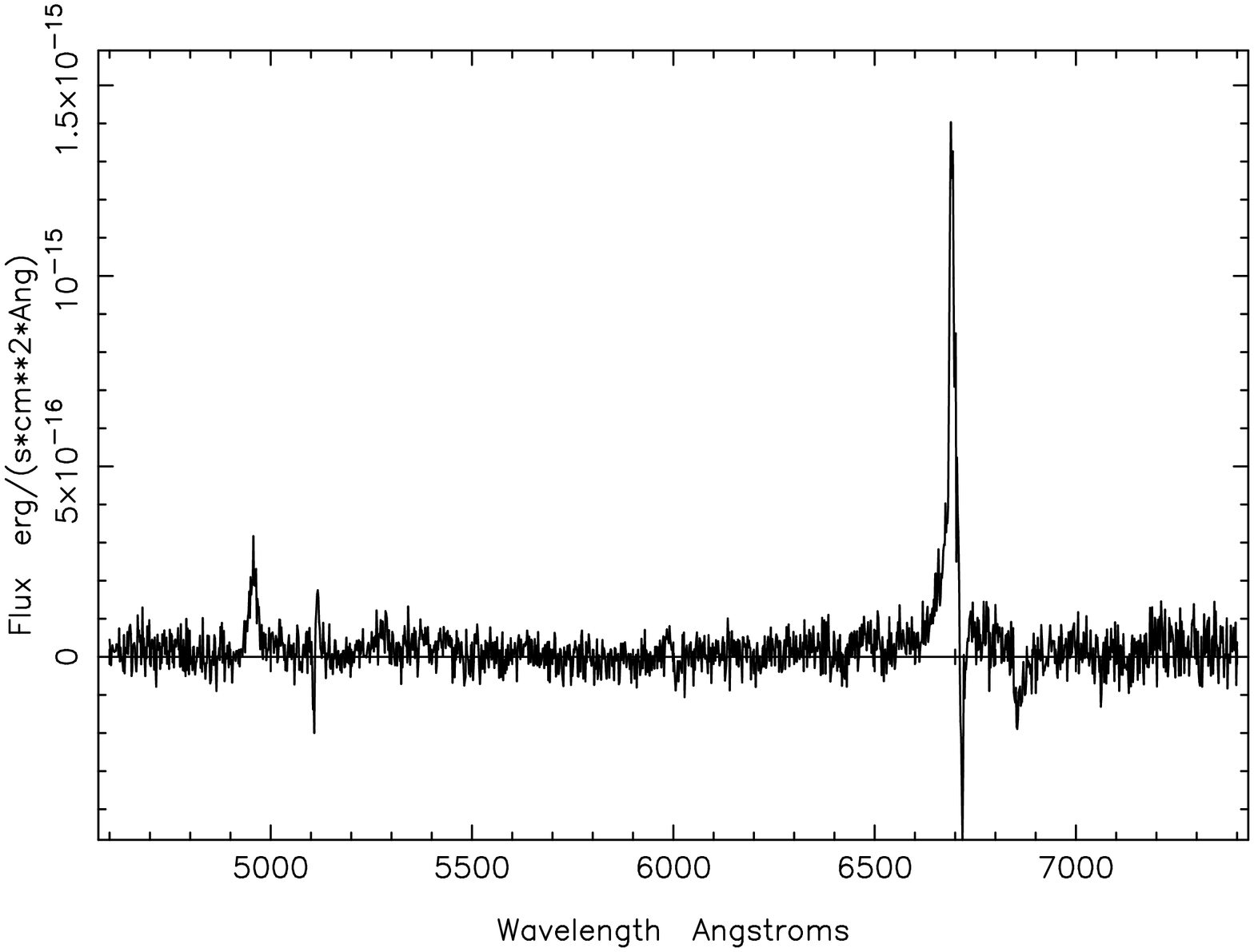,width=4.5in,angle=-90,clip=}}\end{minipage}

\begin{minipage}{\textwidth}{
{\bf Figure 2:} Residual polarised flux after removal of the `smooth' component
in the polarisation.  Data are shown for (a) NGC~2992 and (b) 18325--5926.  The
negative feature near H$\alpha$ in (b) is due to the overcorrection of narrow
[NII].  Broad H$\beta$ is evident in addition to a broad H$\alpha$ component
in both sources.
}\end{minipage}

\newpage

\begin{minipage}{6.5in}{
\psfig{file=Figure3.ps,width=6.5in,angle=0,clip=}}\end{minipage}

\begin{minipage}{\textwidth}{
{\bf Figure 3} (a) Infrared colour-colour plot similar to that presented in
Lumsden et al.\ (2001).  The solid line represents the track of a suitably
reddened Seyfert 1, the dashed line represents the mean for the local starburst
population and the dot-dash line is the limit for mixing between the two types.
We clearly find HBLRs in galaxies with cool colours consistent with a starburst
origin for the infrared emission.  (b) Infrared-radio colour-colour diagram,
similar to that presented in Lumsden et al.\ (2001) and Tran (2001, 2003).
Powerful AGN cluster in the top left of this diagram.  However, we also find
HBLRs in galaxies in the lower right region that Tran describes as being
populated by `pure' Seyfert 2 galaxies.  
}\end{minipage}

\end{document}